\documentclass[aps,prb,twocolumn,showpacs]{revtex4}

\usepackage{graphicx}
\usepackage{dcolumn}
\usepackage{bm}
\usepackage{pifont,wasysym}
\usepackage{pstricks}

\DeclareTextSymbol{\degre}{OT1}{23}
\renewcommand{\vec}[1]{\mbox{\boldmath$#1$}}

\begin{document}

\title{Structure and dynamics of {\it l-}Ge: Neutron scattering experiments and {\it ab initio} molecular dynamics simulations}

\author{Virginie Hugouvieux}
\altaffiliation[Present address: ]{Unit\'e de Recherche Biopolym\`eres, Interactions, Assemblages, INRA, BP 71627, 44316 Nantes cedex 3, France}
\email[Email: ]{virginie.hugouvieux@nantes.inra.fr}
\affiliation{Laboratoire des Collo\"ides, Verres et Nanomat\'eriaux, CNRS, UMR 5587, 
Universit\'e Montpellier 2, 34095 Montpellier Cedex 5, France}
\affiliation{Institut Laue-Langevin, BP 156, 38042 Grenoble cedex 9, France}
\author{Emmanuel Farhi}
\affiliation{Institut Laue-Langevin, BP 156, 38042 Grenoble cedex 9, France}
\author{Fanni Juranyi}
\affiliation{Laboratory for Neutron Scattering, PSI, 5232 Villigen, Switzerland}
\author{Philippe Bourges}
\affiliation{Laboratoire L\'eon Brillouin, CEA Saclay, 91191 Gif sur Yvette cedex, France}
\author{Walter Kob}
\affiliation{Laboratoire des Collo\"ides, Verres et Nanomat\'eriaux, CNRS, UMR 5587, 
Universit\'e Montpellier 2, 34095 Montpellier Cedex 5, France}
\author{Mark R. Johnson}
\affiliation{Institut Laue-Langevin, BP 156, 38042 Grenoble cedex 9, France}

\begin{abstract}
We report the first measurements of the dynamics of liquid germanium ({\it l-}Ge)
by quasi-elastic neutron scattering on time-of-flight and triple-axis
spectrometers. These results are compared with simulation data of
the structure and dynamics of {\it l-}Ge which have been obtained
with {\it ab initio} density functional theory methods. The
simulations accurately reproduce previous results from elastic and
inelastic scattering experiments, as well as the $q$-dependence of the
width of the quasi-elastic signal of the new experimental data. In order to
understand some special features of the structure of the liquid we have
also simulated amorphous Ge. Overall we find that the atomistic model
represents accurately the average structure of real {\it l-}Ge as well as
the time dependent structural fluctuations. The new quasi-elastic neutron scattering data allows us to investigate to what
extent simple theoretical models can be used to describe diffusion in {\it l-}Ge.
\end{abstract}

\maketitle

\section{Introduction}

The investigation of the structure of the condensed phases of germanium
is of importance for fundamental science since this material exhibits polymorphism 
in the solid state and pronounced changes in density and bonding upon melting. Research in this field is also relevant to several
applied physics problems and semiconductor technology.

In its solid phase germanium shows a large degree of polymorphism. In
its stable crystalline phase at ambient pressure, the crystal has a
diamond structure in which each Ge atom is surrounded by four covalently
bonded first neighbors in a tetrahedral formation. This phase of Ge has a very low density with respect to a close-packed
structure and is
semiconducting. Upon application of pressure, the tetrahedrally bonded network
is disrupted and the number of neighbors and the density increase. Around
100 kbar, a phase transition to the metallic $\beta$-tin (or white-tin)
structure occurs and at even higher pressures an hexagonal phase and a
close-packed phase are found\cite{Vorha1986}.

At standard pressure the melting temperature of crystalline Ge
is $T_m$~=~1210.4~K. At the melting transition the tetrahedral
network is disrupted and the
average coordination number increases from 4 to about 7, which is still
low compared to other liquid metals which typically have coordination number between
9 and~12.  Liquid germanium is metallic \cite{Glazov1969}. Germanium can also be
produced in an amorphous phase in which distorted tetrahedra are connected in a continuous random network with defects. Amorphous germanium is semiconducting.

Experimental investigations of the density and bonding changes in different phases of germanium are numerous, using both X-ray
\cite{Isherwood1972,Waseda1975,Filipponi1995} and neutron diffraction
\cite{Gabathuler1979,Davidovic1983,Bellissent-Funel1984,Salmon1988,Kawakita2002}
techniques. The static structure factor shows features that differ from those found in 
simple liquids: The first diffraction maximum is unusually low and has
a shoulder on its high momentum transfer side. Thus a hard-sphere model
does not seem to be appropriate for describing the arrangement of atoms in
{\it l-}Ge and several models have been proposed based on the short-range
order found in white-tin (also called $\beta-$tin)~\cite{Isherwood1972},
on the coexistence of randomly distributed, molten metal-like
atomic arrangement and covalent crystal-like atomic arrangement
\cite{Waseda1975}, on a mixture of fourfold and highly coordinated
metallic arrangements \cite{Gabathuler1979} or on a quasicrystalline
model of {\it l}-Ge \cite{Bellissent-Funel1984}. However none of these
accounts for all of the structural features of {\it l}-Ge.

In order to investigate the dynamics of {\it l-}Ge, Hosokawa and
co-workers \cite{Hosokawa2001} used inelastic X-ray scattering to measure the spectra for
momentum transfers $q$ from 0.2 to 2.8~\AA$^{-1}$. The
phonon dispersion curve extracted from the low $q$ spectra matches
the hydrodynamic sound velocity \cite{Yoshimoto1996}. The study of
the relationship between $S(q)$ and $\Gamma_q$, the half width of
$S(q,\omega)$ at half maximum, shows that at higher $q$ the measured
spectral width cannot be described by the relationship given by de
Gennes for simple dense fluids \cite{Gennes1959} nor the expression
for spectra of dense hard-sphere fluids given by Cohen and co-workers
\cite{Cohen1987}. According to Hosokawa {\it et al.}~\cite{Hosokawa2001},
this discrepancy supports the view that neither the structure
nor the dynamics of {\it l-}Ge can be described using a single pair
interaction. However, Ashcroft~\cite{Ashcroft1990} proposed a cluster
model with transient covalent structures in the liquid which
seems to be consistent with the experimental results. In this model,
the first maximum in $S(q)$ reflects the inter-cluster correlation while
the shoulder on the high $q$ side and the remaining oscillations are
associated with the atom-atom contributions. Regarding the experimental
results on the dynamics, Hosokawa {\it et al.} put forward the idea that the 
narrowing of the quasi-elastic line at the
position of the maximum in $S(q)$, followed by an increase of $\Gamma_q$
at $q$ values that correspond to the location of the shoulder in $S(q)$,
indicates that the covalent structures are diffusing slowly while
individual atoms are subject to rapid translational motion. 

Also on the theoretical side, much effort has been devoted to modeling and
reproducing both structural and dynamic features of {\it l-}Ge.  In 1985,
Stillinger and Weber proposed a model potential function comprising
both two- and three-atom contributions to describe the interactions in
solid and liquid states of Si \cite{Stillinger1985}. Molecular dynamics
simulations of {\it l-}Si using this potential give structural properties
in good agreement with the measurements. Subsequently this potential was also used
for simulating amorphous germanium \cite{Ding1986}.  Later on
the potential devised for {\it l-}Si was used with a different set of
parameters which were obtained by fitting crystalline as well as liquid
Ge phases \cite{Yu1996}, the corresponding static structure factors and
pair correlation functions therefore being in good agreement with the
experimental results. However, it was found that the computed diffusion
coefficient is larger than the experimental values by a factor of 2.

There have also been a number of {\it ab initio} molecular dynamics
(AIMD) simulations of {\it l-}Ge, starting in 1993 with the work of
Kresse and Hafner who found a quite good agreement between simulated
and experimental pair correlation functions and electronic densities of
states \cite{Kresse1993}. Subsequently they extended their study to the
liquid metal-amorphous semiconductor transition \cite{Kresse1994}. Their
results for the pair correlation function and the static structure
factor are in good agreement with experimental measurements for the
amorphous phase. According to these authors, the results for the
liquid indicate a broad homogeneous distribution of local bonding
configurations and show that both classes of models proposed earlier
(mixture of fourfold coordinated and highly coordinated metallic
arrangements \cite{Gabathuler1979} and both sixfold coordinated and
metallic arrangements \cite{Isherwood1972}) are unrealistic. The
simulation of supercooled and amorphous Ge shows the increase of the
local tetrahedral order with decreasing temperature. The self-diffusion
coefficient extracted from the simulated trajectory close to the melting point, $D$~=~1.0$\cdot
10^{-4}$~cm$^2$/s, is in good agreement with the measured value. Similar
results about the structure and diffusion of {\it l-}Ge were found by
Takeuchi and Garz\'on \cite{Takeuchi1994} using the {\it ab initio}
Car-Parrinello molecular-dynamics scheme \cite{Car1985}. A further
study was performed by Godlevsky {\it et al.}~\cite{Godlevsky1995}
which yielded similar results for both the structure factor and the
pair correlation function.  Kulkarni and collaborators performed
similar simulations and studied the influence of the temperature
between 1250~K and 2000~K on the structural properties of the liquid
\cite{Kulkarni1997}. These simulations show that liquid Ge is a good
metal but has some special short-range order arising from residual
covalent bonding. Only at the highest temperatures, {\it l-}Ge seems
to evolve toward a more conventional close-packed liquid metal.
Calculations of the dynamic structure factor from an {\it ab initio} MD simulation were performed by Chai and co-workers \cite{Chai2003}. They
show good agreement with the experimental data, although the statistical
noise due to the small number of atoms (64) is rather large and has to be
convoluted with a Gaussian of width 2.5~meV representing the experimental
resolution function.

As can be seen from these experimental and theoretical studies,
experimental investigations of {\it l-}Ge are presently not able to
determine the microscopic, atomic arrangements in the liquid, but
the experiments provide the data which is needed to validate theoretical models
and simulations.  In this paper, we present a new study of liquid germanium
using both quasi-elastic neutron scattering (QENS) experiments and AIMD
simulations. The QENS technique provides much clearer insight into diffusive
dynamics than inelastic X-ray scattering in an intermediate $q$ range
due to the higher resolution and approximately Gaussian lineshape of the spectrometer resolution function.
The QENS signal is typically most prominent at the $q$-range around
the first sharp diffraction peak. QENS data therefore relates directly to
local structural fluctuations on a length scale which is most relevant
to the on-going discussion about the structure of {\it l-}Ge. Section
\ref{sec:experiment} presents details and results of the QENS
measurements. Section \ref{sec:simulation} is dedicated to the {\it ab
initio} simulations of the structure and dynamics of {\it l}-Ge and to the
comparison with new and existing experimental results and in Section~\ref{sec:conclusion}
we summarize and conclude the present work.

\section{Neutron scattering study of \lowercase{\it l-}G\lowercase{e}}
\label{sec:experiment}

QENS measurements on {\it l-}Ge were performed on
cold neutron, time-of-flight (Focus at the Swiss spallation neutron source SINQ, PSI, Switzerland
\cite{focus_web,Janssen1997,Janssen2000}) and triple-axis (4F1 at Laboratoire L\'eon Brillouin,
France \cite{4F1_web}) spectrometers. Note that due to the high sound velocity,
$c_s$~=~2682~m/s~\cite{Yoshimoto1996}, the acoustic phonon of {\it l-}Ge
could not be measured.

\subsection{Sample preparation and environment}

Solid Ge samples were obtained either as a powder (obtained commercially
with a purity of 99.99\% in weight) or as pure solid pieces (less
sensitive to oxidation and higher packing factor). Solid Ge was inserted
into a cylindrical cell (6~mm in diameter) made of vitreous silica which
was then sealed under vacuum. Sample preparation was performed either
under nitrogen atmosphere or vacuum, in order to avoid contamination
of dioxygen (which would lead to formation of GeO or GeO$_2$ at high
temperatures).  The silica cell was suspended from the end of the
sample stick in a niobium envelope.  The melting temperature of {\it
l}-Ge is $T_m$~=~1210.4~K. During the neutron scattering experiments,
Ge was heated in standard furnaces.  After melting, the height of the
sample was between 2 and 5~cm, depending on the initial solid form of Ge.

\subsection{Experimental setup and data analysis}

In both experiments, the sample temperature was comprised between 1260~K and
1400~K, with a sample of 6~mm in diameter.

On Focus, the measurements were performed at incident wavelengths of
3~\AA~and 4~\AA. Momentum transfer ranges from 0.35 to 3.8~\AA$^{-1}$ at
$\lambda_i$~=~3~\AA. Background measurements with comparable statistics were performed at both wavelengths
for the furnace, empty sample cell and niobium envelope. The instrument
resolution function was determined using a cylindrical vanadium sample
of diameter 10~mm.  The elastic energy resolution at incident wavelength
of 3~\AA~and 4~\AA~ was of the order of 500~$\mu$eV and
200~$\mu$eV, respectively. The data were processed with standard software at the
instrument (NATHAN).

On 4F1 constant $q$ scans were performed between
$q$~=~0.3~\AA$^{-1}$ and $q$~=~3.9~\AA$^{-1}$, although the QENS signal
could only be measured unambiguously for $q >$ ~1.5 \AA$^{-1}$. The data
were processed with the standard 4F1 program. 

\subsection{Experimental results} 

\begin{figure}
\includegraphics*[width=8.6cm]{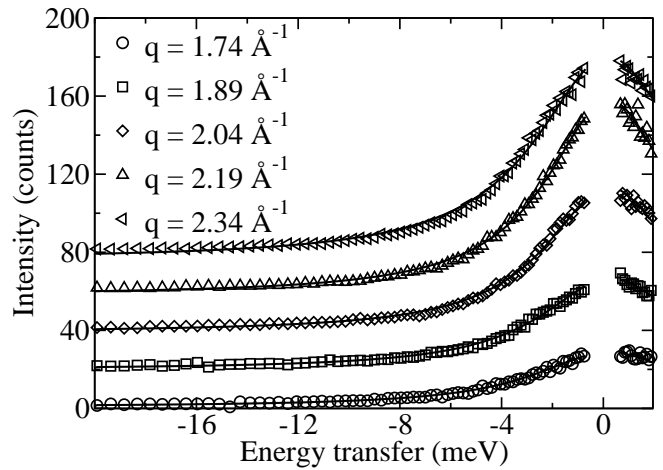}
\caption{Time-of-flight measurements of $S(q,\omega)$ for {\it l-}Ge
at different $q$ values ($\lambda_i$~=~3~\AA; $T$~=~1260~K); different spectra are
shifted vertically by 20 counts. The solid lines show the fitted Lorentzians.}
\label{fig:fitted_spectra}
\end{figure}

\begin{figure}
\includegraphics*[width=8.6cm]{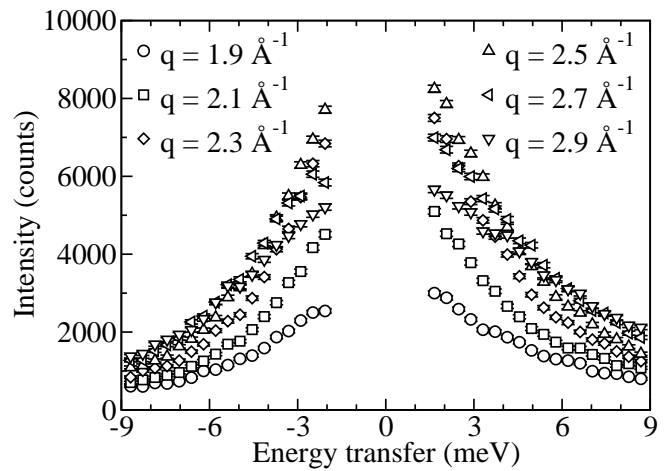}
\caption{Triple-axis measurements of $S(q,\omega)$ for {\it l-}Ge at
different $q$ values around the position of the first maximum in $S(q)$
($k_f$~=~2.662 \AA$^{-1}$, $T$~=~1300~K). }
\label{fig:scans_4F1}
\end{figure}

The background signal was almost as strong as the signal from
{\it l-}Ge. A slight misalignment between the sample and empty
cell measurements meant that the data around the elastic peak had to be
discarded. The difference spectra were fitted using a single Lorentzian
for each $q$ (see Fig. \ref{fig:fitted_spectra} for Focus results). The
QENS intensity increases and the width decreases as $q$ approaches
2.5~\AA$^{-1}$. The triple-axis measurements give similar results (see
Fig. \ref{fig:scans_4F1}). The evolution of the QENS signal with $q$
will be discussed below with the results from the simulation.

\section{{\it Ab initio} simulation study of \lowercase{\it l-}G\lowercase{e}} 
\label{sec:simulation}

\subsection{Simulation procedure}

All calculations were performed with the Vienna Ab initio Simulation
Package (VASP) \cite{Kresse1996, Kresse1996a}, which is based on
density functional theory.  We carried out most simulations of {\it l-}Ge
for an ensemble of 64 atoms in a cubic supercell with periodic boundary
conditions. The simulated density was $\rho=5.53~g/cm^3$, which corresponds to
a box of length 11.175~\AA.

The atomic dynamics was described in the microcanonical ($N$,$V$,$E$)
ensemble. Ultrasoft Vanderbilt pseudopotentials \cite{Vanderbilt1990}
and the generalized gradient approximation (GGA) with the functional
PW91 for the exchange-correlation energy were used.  The Brillouin zone
was sampled by a $3\times3\times3$ Monkhorst-Pack mesh of $k$-points
\cite{Monkhorst1976}.  The wave functions were expanded in a basis of more
than 32000 plane waves, corresponding to an energy cut-off of 139.2~eV.
One MD run was performed in a cell containing 200 atoms for which the $k$-point
grid contained only the gamma point.

The starting configurations for the {\it ab initio} simulations were
generated from classical MD simulations using the Stillinger and Weber
potential \cite{Stillinger1985}. For this purpose, systems of 64 Ge
atoms were equilibrated at the desired temperatures (three independent
samples were extracted from a 50~ps run at high temperature (1700~K);
these samples were then equilibrated at the desired temperature by velocity
scaling during 10~ps).

After switching to the {\it ab initio} ($N$,$V$,$E$) MD, and an equilibration time of 5~ps, three runs
of 30~ps using a time step of 3~fs were performed at three different
temperatures corresponding to the liquid (1390~K, 1080~K) and supercooled
(680~K) states, all at the same density.  A similar simulation scheme was used to perform a liquid MD run at 1390~K for the cell containing 200 atoms, the 
NVE trajectory spanning 15~ps. Three independent
configurations of the trajectory at 680~K were quenched at a cooling
rate of 68~K/ps in order to obtain three amorphous configurations of
the system.

\subsection{Numerical results} \label{sec:AIMD_results}

The mean-squared displacement (MSD) of atoms is one of the dynamic
quantities that allows to check whether the system is a liquid or not.
The MSD is computed from the relation

\begin{equation}
\delta r^2(t) = \langle [ \vec{r}_i(t) - \vec{r}_i(0)]^2 \rangle  \quad ,
\end{equation}

\noindent
and allows the diffusion constant $D$ to be determined via the Einstein relation :

\begin{equation}\label{eq:MSDlinear}
D = \lim_{t\to \infty} \frac{\delta r^2(t)}{6t} \quad .
\end{equation}

\noindent

\begin{figure}
\includegraphics*[width=8.6cm]{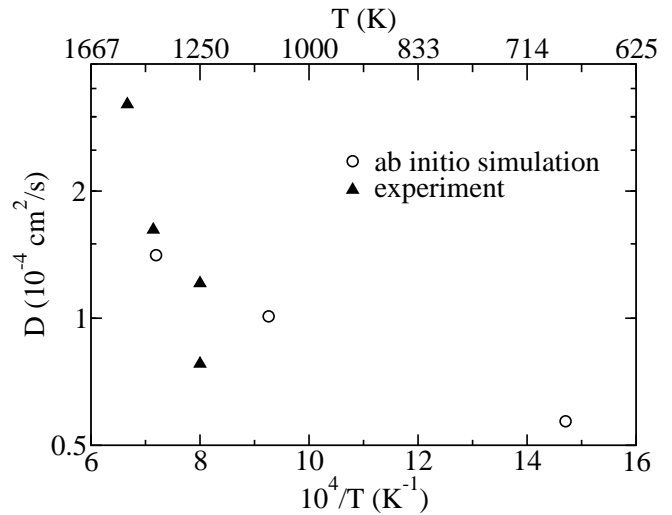}
\caption{Arrhenius plot of calculated and experimental diffusion
constant $D$ (experimental data from Ref. \onlinecite{Pavlov1970}). Note that the vertical scale is logarithmic.}
\label{fig:Dexpsimul}
\end{figure}

In Fig.~\ref{fig:Dexpsimul} we compare the values for $D$ as
obtained from the simulations with experimental data from Pavlov and
Dobrokhotov~\cite{Pavlov1970}. For $T$ around 1400~K, very similar values of $D$
are obtained from experiment and {\it ab initio} MD, while according
to Ref. \onlinecite{Yu1996} classical MD overestimates $D$ by a factor
of~2, which indicates that from this point of view the {\it ab initio}
calculation is indeed more accurate than simulations with classical
force fields. Having performed liquid simulations at 3 temperatures, we are able to observe an Arrhenius behavior of the diffusion constant with an
activation energy around 0.72~eV. This quantity corresponds to the average energy needed by a Ge atom to escape from its local environment and undergo diffusion and it is characteristic of a strong liquid.

\begin{figure}[!htbp]
\includegraphics*[width=8.6cm]{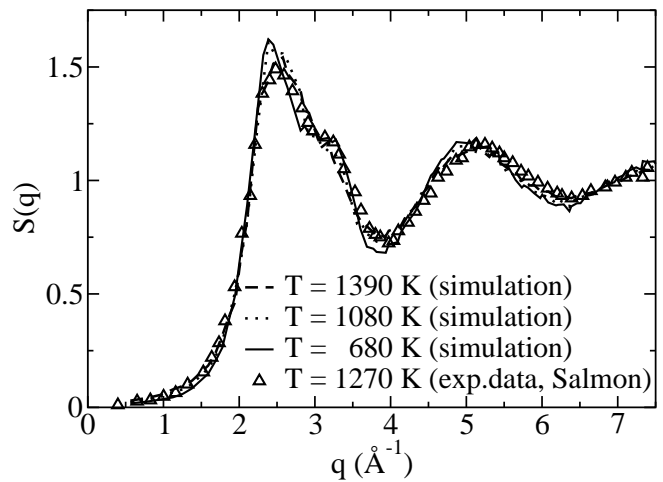}
\caption{Static structure factor of {\it l-}Ge: Simulated data
for $T$~=~1390~K (dashed line), $T$~=~1080~K (dotted line), and $T$~=~680~K (full
line) and experimental data by Salmon \cite{Salmon1988} at $T$~=~1270~K
(triangles).}
\label{fig:Sq_Ge_VASP}
\end{figure} 

\begin{figure}[!htbp]
\includegraphics*[width=8.6cm]{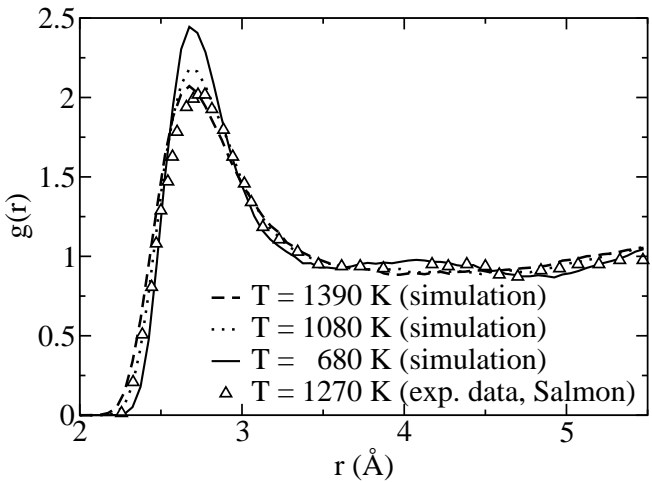}
\caption{Pair correlation function of {\it l-}Ge: Simulated data: $T$~=~1390~K (dashed
line); $T$~=~1080~K (dotted line) and $T$~=~680~K (full line) and neutron scattering 
data by Salmon~\cite{Salmon1988} at $T$~=~1270~K (triangles).}
\label{fig:gr_Ge_VASP}
\end{figure} 

The structure of the liquid can be characterized by the 
static structure factor which is defined as~\cite{Hansen2006}:

\begin{equation} \label{eq:sq_density}
S(\vec{q})= \frac{1}{N} \sum_{j=1}^N \sum_{l=1}^N \langle \exp(i \vec{q} \cdot (\vec{r}_j - \vec{r}_l)) \rangle .
\end{equation}

\noindent
It was computed using, at each temperature, 10000 atomistic configurations. 
The average is also performed over all $\vec{q}$ vectors of
the same magnitude. The results are shown in Fig. \ref{fig:Sq_Ge_VASP}
together with the experimental result by Salmon \cite{Salmon1988}.

At $T$~=~1390~K the main sharp peak occurs at
$q$~=~2.53~\AA$^{-1}$ which is very close to
the experimental values ($q_{max} \approx$~2.50~\AA$^{-1}$)
\cite{Bellissent-Funel1984,Salmon1988,Waseda1975}. At lower
temperatures, the peak is shifted to slightly lower $q$ values, which is
consistent with measurements in the temperature range from 1270~K to
1820~K \cite{Kawakita2002}.
The characteristic feature of {\it l-}Ge shown by neutron
diffraction studies \cite{Kawakita2002}, the distinct shoulder at
$q$~=~3.27~\AA$^{-1}$, i.e. on the high-$q$ side of the first sharp
peak, is reproduced well by the simulation. As $T$ decreases, the height of the main
peak increases and the shoulder becomes more distinct. These two
features were described at higher temperatures in the simulation work
by Kulkarni {\it et al.} who found that at $T$~=~2000~K the shoulder
disappears~\cite{Kulkarni1997}. The second peak of the static structure factor occurs at 5~\AA$^{-1}$
which is slightly shifted from the experimental value reported by
Salmon \cite{Salmon1988} who found $q=5.1$~\AA$^{-1}$ (see also Waseda \cite{Waseda1975}),
although the experimental value seems to have an uncertainty of
$\approx$~0.1~\AA$^{-1}$~~\cite{Isherwood1972,Orton1973}.

\begin{figure}[!hbtp]
\includegraphics*[width=8.6cm]{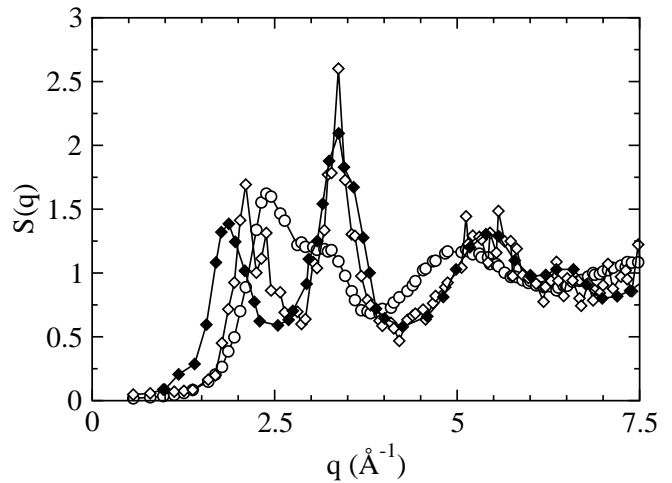}
\caption{$S(q)$ for supercooled and amorphous Ge:
Comparison between simulation (supercooled state at $T$~=~680~K (open
circles) and amorphous state (open diamonds)) and experimental data, from
Ref. \onlinecite{Etherington1982}, as obtained from neutron scattering with an amorphous sample (filled diamonds).}
\label{fig:sq_amorph}
\end{figure}

The pair correlation function $g(r)$, defined as~\cite{Hansen2006}

\begin{equation}
\rho g(\vec{r}) = \frac{1}{N} \sum_{i=1}^N \sum_{j \neq i}^N
\langle \delta(\vec{r} - (\vec{r}_i - \vec{r}_j)) \rangle \quad ,
\end{equation}

\noindent
allows the structure of the system in real space to be characterised. A
comparison of simulated and experimental $g(r)$ is given in Fig.~\ref{fig:gr_Ge_VASP}. In particular we see that $g(r)$ hardly shows a
second peak at distances beyond the first peak, in marked contrast to
the situation found in simple liquids such as hard spheres. The structure of {\it l-}Ge is significantly different
from that of simple liquids. In addition
the height of the main peak increases with decreasing temperature but the
structure at larger distances remains basically unchanged. Indeed both $S(q)$ and $g(r)$ show that structural changes over 
a wide range of temperatures from 680 to 1390 K are generally rather small.

\begin{figure*}[!hbtp]
\includegraphics*[width=17.2cm]{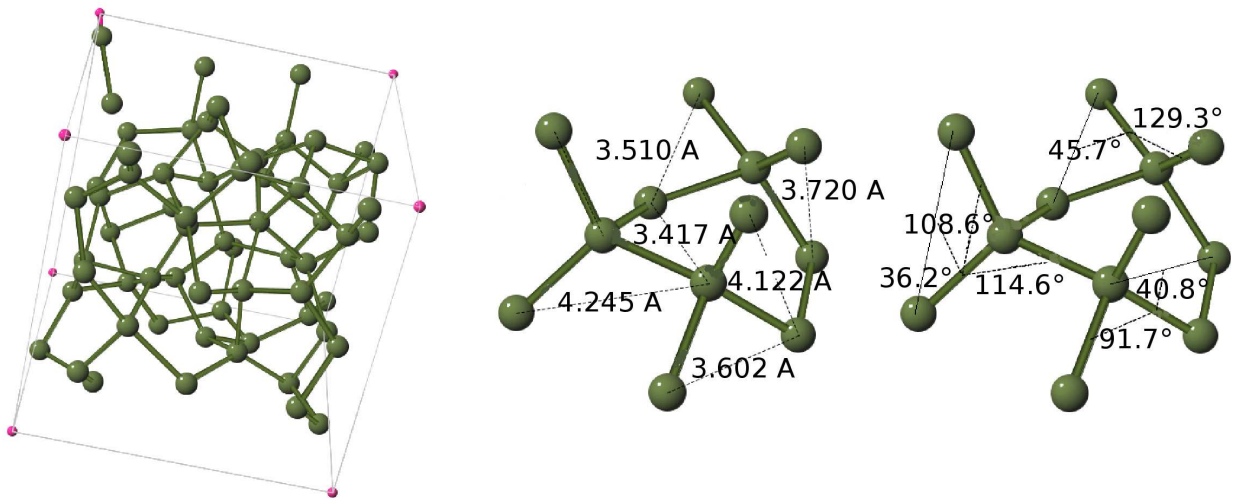}
\caption{(Color online) Calculated structure of amorphous Ge at $T$~=~0~K. Left:
Snapshot of the whole simulation box (with small spheres representing the
corners of the box). Middle: Interatomic distances. Right: Bond angles.}
\label{fig:snapshot_aGe}
\end{figure*}

Structural changes become more significant if the liquid phase is quenched to produce an amorphous phase. This
can be seen in Fig.~\ref{fig:sq_amorph} 
in which we show $S(q)$ for the supercooled liquid
at 680~K and the amorphous state.  The amorphous phase was obtained by
cooling down, over 10 ps and at constant volume, three independent configurations of the supercooled state
at 680~K to $T=0$~K. Thus for this state $S(q)$ was
calculated by averaging over only three configurations, which is the
reason for the substantial noise in the simulation data.  
In $S(q)$, the first peak is shifted to higher $q$
reflecting the high, liquid density that we have imposed on the amorphous structure. 
At higher $q$, the agreement between experiment and simulation is much better.
Looking at snapshots of the system
in the amorphous state, one clearly sees a tetrahedral structure in which
a Ge atom occupies the center of a tetrahedron. The calculated $g(r)$ for the amorphous state (not shown here) shows a main peak shifted to lower interatomic distances (as compared to the liquid state) while the intermediate peak near 4~\AA~becomes much more pronounced, in agreement with experimental findings \cite{Etherington1982}. The main peak in $g(r)$
corresponds to the interatomic distance between the center and a vertex
of the tetrahedron while the higher interatomic distance at 4~\AA~is
associated with the distance between two vertices of the tetrahedron
(see also Fig. \ref{fig:snapshot_aGe}). Despite imposing the higher liquid density on our amorphous structures, the 
shortest Ge-Ge distances are overestimated by several 0.1~\AA.

\begin{figure}[!hbtp]
\includegraphics*[width=8.6cm]{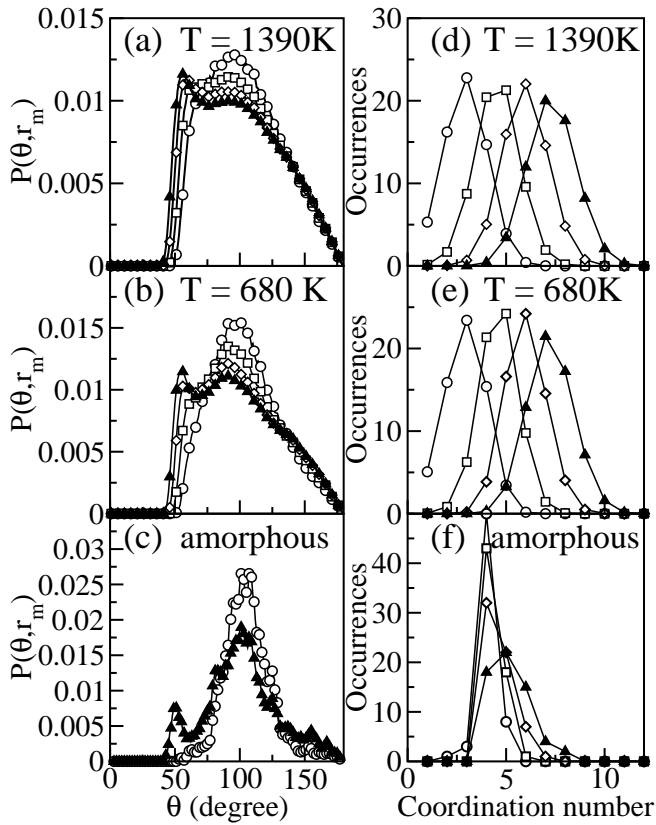}
\caption{Angular distribution (left) and coordination number (right) for
liquid (top), supercooled (middle) and amorphous (bottom) phases of 
Ge for $r_m$ in the vicinity of the first minimum of $g(r)$. Circles:
$r_m$~=~2.8~\AA; Squares: $r_m$~=~3.0~\AA; Diamonds: $r_m$~=~3.2~\AA;
Triangles $r_m$~=~3.4~\AA. The quantities for the liquid, supercooled,
and amorphous phases are, respectively, averaged over 9000, 6300 and
3 configurations.}
\label{fig:angular}
\end{figure} 

Given the nature of the many-body interactions in germanium and the
resulting open structure, it is interesting to compute the angular
distribution of atoms. For this we define $P(\theta,r_m)$ as the
normalized distribution of angles defined by vectors going from a
reference atom $i$ to two neighbors of $i$ which are within a radius $r_m$
of the reference atom. In Fig.~\ref{fig:angular} we show $P(\theta,r_m)$
for several values of $r_m$ around the minimum in $g(r)$. In the
liquid this minimum is not very pronounced, see Fig.~\ref{fig:gr_Ge_VASP},
leading to an uncertainty regarding the definition of the nearest neighbor
shell. In contrast, in the amorphous state, the
first minimum in $g(r)$ (calculated, not shown here) is clearly defined 
allowing an unambiguous definition of nearest neighbors. We note that any bond angle distribution plot will tend to have a maximum close to 90\degre and that changes in the distribution as a function of radius and temperature are therefore particularly meaningful.

For liquid and supercooled systems, we observe a broad peak around
$\theta$~=~98\degre if $r_m$~=~2.8~\AA. With increasing $r_m$, this
peak tends to shift to smaller angles, around $\theta$~=~90\degre. For
the amorphous phase, the corresponding peak is observed at slightly
larger angles, around 105\degre for $r_m$~=~2.8~\AA, indicative of a slightly more
regular tetrahedral environment in the amorphous phase, even at higher pressure 
as imposed by the liquid density. With increasing
$r_m$, a new peak appears for angles between 55\degre and 60\degre in
the liquid and supercooled phases. At $T$~=~680~K, this feature is more
pronounced than at higher temperatures. In the amorphous phase, this peak
appears at smaller angles, about 50\degre.  A bond angle of 60\degre
is characteristic of metallic bonding, since this value represents the
locally most closely packed structure of the system. The broad peak
around 100\degre is typical of flattened tetrahedra, the symmetric
tetrahedral angle being 109.5\degre, see also Fig.~\ref{fig:snapshot_aGe}.

Regarding the coordination number, most atoms have either 4 or
5 nearest neighbors in the amorphous phase (see right panels
of Fig.~\ref{fig:angular}). On the other hand, in the liquid and
supercooled phases, for typical values of the first minimum in $g(r)$,
3.4~\AA, the coordination number is between 6 and 7.

\begin{figure*}[!hbtp]
\includegraphics*[height=6cm]{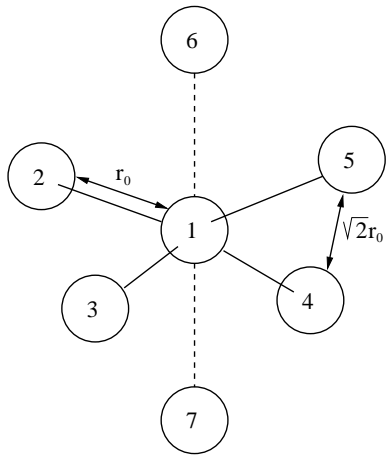}
\caption{Schematic representation of the $\beta$-tin configuration. Atom 1 is at the center of the flattened  tetrahedron formed by atoms 2, 3, 4, and 5. Atoms 6 and 7 are the second nearest neighbors. Distances are: $d_{12}=d_{13}=d_{14}=d_{15}\approx r_0$ ; $d_{23} = d_{34}= d_{45} = d_{52} \approx \sqrt{2} r_0$ ; $d_{24}=d_{35} \approx 2 r_0$.}
\label{fig:beta_tin}
\end{figure*}

We note that obviously the coordination number increases with increasing
radius of the coordination sphere for both liquid and supercooled
systems, which is consistent with the so-called $\beta$-tin hypothesis
\cite{Isherwood1972}. This hypothesis states that in {\it l-}Ge, each
atom has four nearest neighbors on a flattened tetrahedron and two other
neighbors slightly further away on the normal to the plane of
the flattened tetrahedron (see Fig. \ref{fig:beta_tin} for a schematic
representation of the local environment of $\beta-$tin). Denoting by $r_0$
the main interatomic distance reported from $g(r)$, i.e. $r_0\simeq
2.5$~\AA, the $\beta$-tin configuration would lead to interatomic
distances of the order of $r_0$ (between the center and a vertex of
the tetrahedron), $\sqrt{2}r_0 \simeq 3.53$~\AA~(between two adjacent
vertices of the tetrahedron), and $2 r_0 \simeq 5$~\AA~(between
two opposite vertices of the tetrahedron) together with distances
slightly higher than $r_0$ associated with the two further neighbors.

Snapshots of the simulated structure of amorphous Ge are shown in
Fig. \ref{fig:snapshot_aGe}. On the left, the 64 atoms are represented
together with the corners of the simulation box. The bonds between
atoms correspond to two covalent radii (1.22~\AA~for Ge) plus a bond
tolerance of 0.4~\AA~so a bond is drawn for interatomic distances
smaller than 2.84~\AA. The two other plots show a zoom into this
structure. Three ``tetrahedra'' are represented together with some
interatomic distances and bond angles. The covalent bond length of Ge
corresponds to the main peak in $g(r)$ around 2.5~\AA. On the other hand,
the distance between two neighboring vertices of a tetrahedron ranges from 3.5~\AA~to
4.5~\AA, which accounts for the broad peak around 4~\AA~in $g(r)$ (calculated, not shown here). Looking at the bond angles, their values are
mainly distributed between 85\degre and 130\degre with an average value of
105\degre, close to the tetrahedral value (see Fig. \ref{fig:angular}). At
higher interatomic distances (see distribution for $r_m$~=~3.4~\AA~in
Fig. \ref{fig:angular}), another peak appears around 50\degre. This one
is due to the angles formed by the two following vectors: The first one
goes from a vertex to the center of the tetrahedron while the second
one goes from the same initial vertex to another vertex of the same
tetrahedron. This is also shown in Fig. \ref{fig:snapshot_aGe}.

\begin{figure}[hbtp]
\includegraphics*[width=8.6cm]{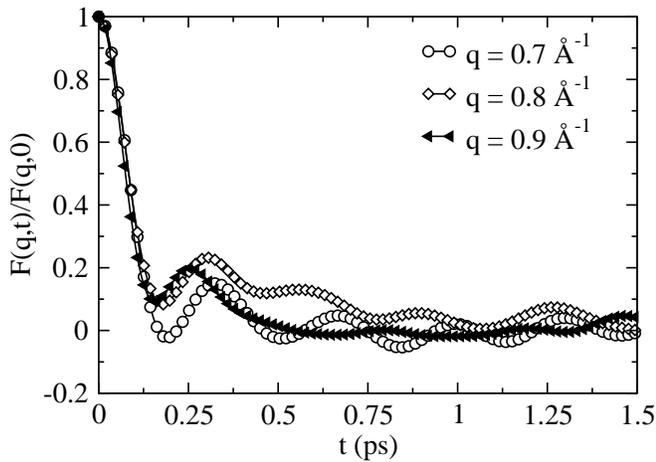}
\caption{Collective dynamic properties of {\it l-}Ge at low $q$ values:
Simulated intermediate scattering function $F(q,t)$ at $T$~=~1390~K
normalized by its value at $t=0$.}
\label{fig:Fqt_Sqw}
\end{figure}

Having established that the liquid simulations do indeed describe well
the corresponding experimental system and having studied the structural
aspects, we now investigate the dynamic features of the system. For
this we start with the intermediate scattering function which is defined
as~\cite{Hansen2006}

\begin{equation}
F(\vec{q},t) = \frac{1}{N} \sum_{j=1}^N \sum_{k=1}^N \langle \exp[i \vec{q} \cdot
(\vec{r}_j(t) - \vec{r}_k(0))] \rangle \quad .
\end{equation}

\noindent
Since the system is isotropic, we have performed an average over the different
orientations of $\vec{q}$.

Figure \ref{fig:Fqt_Sqw} shows the (normalized) $F(q,t)$ of {\it l-}Ge at
1390~K and for low $q$ values from the large simulation cell. 
The oscillations are associated with the
presence of the acoustic mode at these wave-vectors. The period of these
oscillations decreases with increasing $q$ and also they become more
damped, which is directly associated with the shape of the dispersion
and damping of the acoustic phonon as a function of $q$, as will be
discussed in more detail below.

\begin{figure}[!h]
\includegraphics*[width=8.6cm]{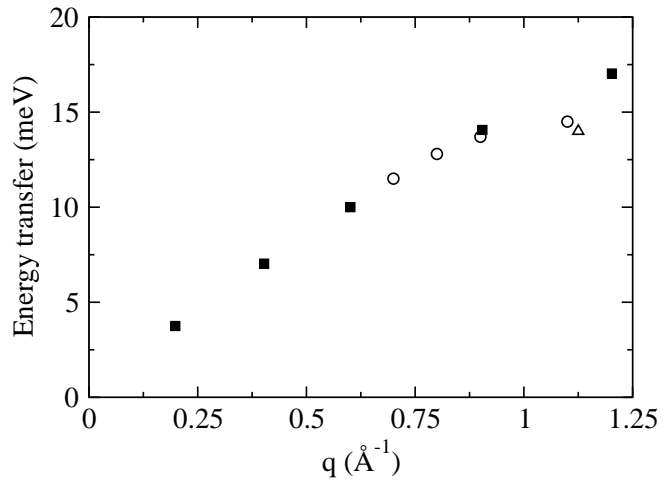} 
\caption{Dispersion of the frequency of the acoustic phonon as a function
of $q$: Comparison between inelastic X-ray scattering data at $T$~=~1250~K
from  Ref.~\onlinecite{Hosokawa2001} (squares) and simulation at $T$~=~1390~K 
(circles for the 200 atom cell, triangles for the 64 atom cell).}
\label{fig:comp_disp_VASP}
\end{figure}

The dynamic structure factor $S(q,\omega)$ was computed by a Fourier
transform in time of $F(q,t)$~\cite{Hansen2006}. In a periodic system of cell side $L$ only the reciprocal lattice 
points $(n_x,n_y,n_z).2\pi/L$ are accessible. The first q-point corresponds to a spatial correlation over a distance $L$, which is not meaningful in a cell of side $L$. $q$-points bigger than $4\pi/L$ correspond to correlations over distances 
shorter than $L/2$, which are uniquely defined. In the intermediate range there is a progressive increase in meaningful correlations which correspond to distances between atoms close to the centre and those close to the 
apexes of the cell. The lowest unambiguous $q$-values, calculated in 0.1~\AA$^{-1}$ $q$-strips, for the 200 atom cell start at 
0.7~ \AA$^{-1}$ and at 1.1~\AA$^{-1}$ for the 64 atom cell. 
At smaller $q$, $q \leq 1$~\AA$^{-1}$, $S(q,\omega)$ shows a well defined
acoustic mode, the position of which can be compared directly with the
corresponding experimental results from the inelastic X-ray scattering
of Ref.~\onlinecite{Hosokawa2001}. Such a comparison is made in
Fig.~\ref{fig:comp_disp_VASP} and it shows a very good agreement between
experiment and simulations. The width of the acoustic phonons increases with frequency, both in the experiment and 
in our simulations. At 1.2~\AA$^{-1}$ the phonons are hardly visible in the experimental spectrum and in the simulations, 
the width is of the same order as the frequency. Accordingly we estimate the error in the frequency to be a few 
meV. We also mention that the dispersion of the
acoustic mode does not show any significant variation with $T$ in the
temperature range investigated.

\begin{figure}
\includegraphics*[width=8.6cm]{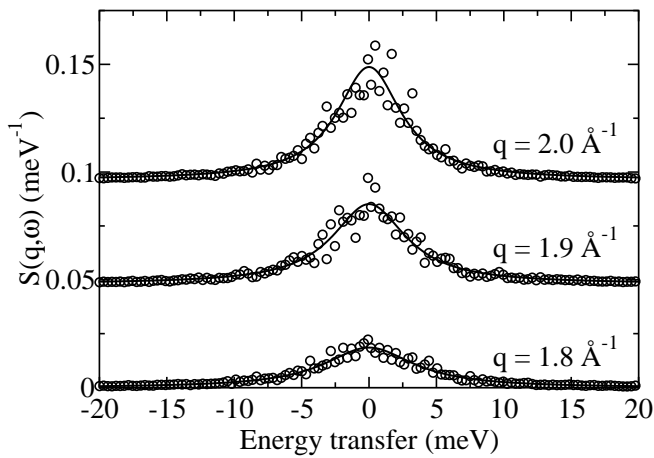}
\caption{Simulated dynamic structure factor for {\it l-}Ge at 1390~K for
different $q$ values (Symbols: Simulation; Line: Fit with a Lorentzian).}
\label{fig:fitting_curves}
\end{figure}

\begin{figure}
\includegraphics*[width=8.6cm]{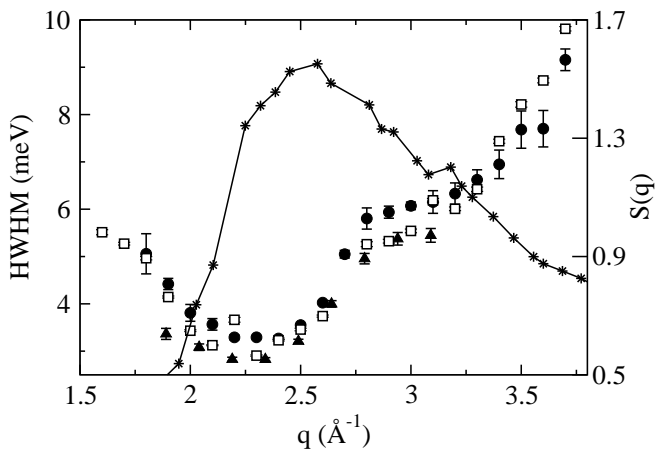}
\caption{Half width at half maximum of $S(q,\omega)$ as a function of $q$ together with the static structure factor. Circles: Triple-axis measurements (4F1); Triangles: Time-of-flight measurements (Focus); Squares: Simulation; Stars and straight line: Simulated $S(q)$ at 1390~K.}
\label{fig:comp_width_simul}
\end{figure}

At higher $q$ values, $q > 1$~\AA$^{-1}$, the QENS signal dominates
$S(q,\omega)$ (see Fig. \ref{fig:fitting_curves}). Fitting the
quasi-elastic line of $S(q,\omega)$ with a Lorentzian allows to
determine its half-width at half maximum (HWHM), the $q$-dependence
of which is shown in Fig.~\ref{fig:comp_width_simul} together with the corresponding structure factor. Unconstrained
fits of the Focus data lead to an underestimation of the spectral
width at low $q$ ($<$~1.8~\AA$^{-1}$), where the QENS intensity is weak,
and also at higher $q$ ($>$~3.2~\AA$^{-1}$), where only the negative energy
transfer part of the spectrum is present, and therefore we show the Focus data only for
1.8~\AA$^{-1}$ $\leq q \leq$~3.2~\AA$^{-1}$. The simulated widths are
obtained from the simulation at 1390~K by fitting $S(q,\omega)$ with
a Lorentzian. From Fig.~\ref{fig:comp_width_simul} we recognize that
the data from the simulation agrees very well with the one from the
experiment. In particular we see that in the simulation as well as in
the  experiments a narrowing of the quasi-elastic peak is observed for
$q$~=~2.4~\AA$^{-1}$ and between $q$~=~2.8~\AA$^{-1}$ and 3.2~\AA$^{-1}$,
which corresponds to the main peak and the shoulder in the static
structure factor, respectively (see line in Fig.~\ref{fig:comp_width_simul}). 
This behavior can be interpreted as de
Gennes narrowing~\cite{Gennes1959}, i.e. the fact that if one considers
the dynamics on length-scales that are close to the one of the local
structure of the liquid, the structure is basically preserved and thus
there is only a relatively small loss of memory and consequently a small
broadening of the dynamic structure factor.

The $q$-dependence of the width of the quasi-elastic peak can be
compared with theoretical predictions based on different models. The
simplest expression for the HWHM for a dense monoatomic liquid is
that the half-width $\omega_q$ is the square root of the normalized second frequency
moment~\cite{Hansen2006}:

\begin{equation} \label{eq:w2_coh} 
\langle \omega_q^2\rangle = \frac{k_B T}{m S(q)} q^2   \quad .
\end{equation} 

In a somewhat refined approach, Cohen {\it et al.}~\cite{Cohen1987}
predicted for a dense hard-sphere fluid the half-width $\omega_h$ to be
given by

\begin{equation} \label{eq:wh}
\omega_h(q) = \frac{D_E q^2}{S(q)} d(q), 
\end{equation}

\noindent
where
\begin{equation}
d(q) = [1 - j_0(q \sigma) + 2 j_2(q \sigma)]^{-1}.
\end{equation} 

\noindent
Here $D_E$ is the self-diffusion coefficient of the hard-sphere fluid
within the Enskog theory, $\sigma$ is the diameter of the hard spheres,
and $j_0(x)$ and $j_2(x)$ are the zeroth- and second-order spherical
Bessel functions, respectively. Note that equation (\ref{eq:wh}) is
only valid for intermediate values of $q$, $1 < q \sigma < \sigma/l$,
where $\sigma/l \gg 1$ with $l$ the mean free path of a particle which,
since we are considering a dense system, is much smaller than $\sigma$.
Approximating the Ge liquid by a system of hard spheres, we have
calculated the quantity $\omega_h(q)$ by using the static structure factor
from the AIMD simulations at 680~K and 1390~K. For the size of the hard
spheres, the location of the first peak in the pair correlation function
gives, see Fig.~\ref{fig:gr_Ge_VASP}, $\sigma$~=~2.8~\AA~at $T$~=~1390~K
and $\sigma$~=~2.7~\AA~for $T$~=~680~K.  For $D_E$, we used the value
of the self-diffusion constant $D$ calculated from the simulation (its
dependence on the mass and the density can be expected to be very similar
to that of the self-diffusion constant of the Enskog theory).

\begin{figure}
\includegraphics*[width=8.6cm]{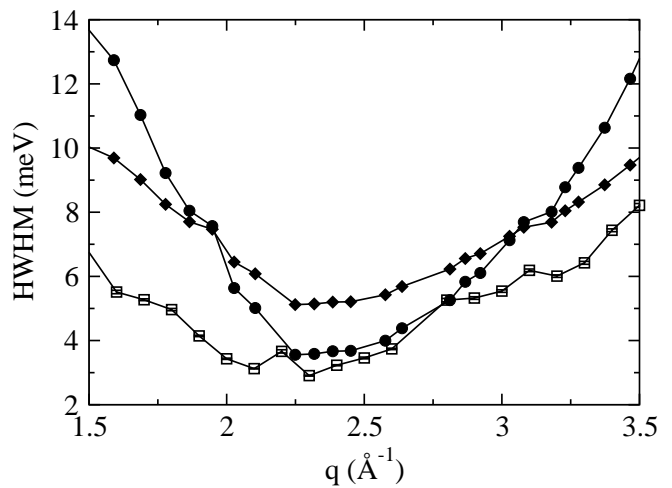}
\caption{Half width at half maximum of the dynamic structure factor as
a function of $q$ for $T$~=~1390~K. Squares: Simulation;  Diamonds:
Second frequency moment $\omega_q$ of a monomatomic liquid, see
Eq.~(\ref{eq:w2_coh}); Circles: $\omega_h$ as calculated from the theory
of dense hard-sphere fluids, see Eq.~(\ref{eq:wh}).}
\label{fig:comp_width_simul_theo}
\end{figure}

These models are compared with the width of the quasi-elastic signal
obtained from AIMD in Fig. \ref{fig:comp_width_simul_theo}.  Both models
show de Gennes narrowing at $q =$~2.3~\AA$^{-1}$. For the dense monoatomic
fluid model, the $q$-dependence of $\omega_q$ is reasonably good but the
value of $\omega_q$ is generally overestimated by a factor of ~2. Such a
systematic discrepancy has also been found for a variety of {\it simple}
liquids (Ar, Kr, Lennard-Jones)~\cite{Cohen1987} and thus it is not
surprising that in the present system, which has a much more complex
structure, we find a similar deviation. In contrast to this the width
$\omega_h$ predicted by the hard sphere model is in better agreement close
to $q=2.3$~\AA$^{-1}$ but the increase in width away from this $q$-value
is over-estimated. Also this result is in qualitative agreement with the
ones reported by Cohen {\it et al.} for simple liquids~\cite{Cohen1987}
which thus gives evidence that Eq.~\ref{eq:wh} is not valid for
simple liquids and for network forming systems.

\section{Discussion and Conclusion}
\label{sec:conclusion}

Liquid Ge is a system of fundamental interest for its structure and fluctuations.
Accordingly a number of models have
been proposed to describe the local structure of this liquid. Due to the
complexity of the system, models can only be validated by comparison with
experimental data but it is found that static structure measurements presently available are
not sufficient to distinguish between various plausible models. Since
the local structure of a liquid is time dependent, a more stringent
requirement of any model is that the structural fluctuations should also
be correctly reproduced.  In this context,
we have presented new QENS measurements of the diffusive dynamics of Ge
on the length scale of $\sim$3~\AA, which is most pertinent to the structural
issues on the scale of the nearest neighbor distance.

To understand this new QENS data, we have also performed {\it ab initio}
simulations of liquid and amorphous Ge, the latter highlighting features
of the liquid phase. These simulations not only reproduce accurately the
quasi-elastic broadening but also IXS data of acoustic phonons at small $q$, 
which concerns the structure on a length scale $>$ 6~\AA.
In addition, we have found very good agreement for the static structure factor, the pair correlation function and
the diffusion constant, the latter enabling an activation energy of 0.7~eV to be determined for diffusion 
processes, which is characteristic of a strong liquid.

 The success of AIMD is
in contrast to classical MD simulations that use two and three body
potentials that, although they reproduce reasonably well the average
structural properties, do not give an accurate description of the
dynamics~\cite{Ding1986,Hugouvieux2004a}. The QENS data and simulations
also allow theoretical models, which relate the spectral line width to
the static structure factor, to be tested. We have found that the second
frequency moment given by Eq.~(\ref{eq:w2_coh}) does not give a good
description of the line width in the whole $q$-range whereas expression
(\ref{eq:wh}) describes this width well around the main peak in the
static structure factor.

Simulations also allow models to be investigated at the atomic level. Analysing
the AIMD trajectories shows an average coordination number of 6-7 for
the liquid and a broad range of bond angles with an average value of
~100 degrees. In contrast to this, for amorphous Ge, the coordination
number is close to 4 and the average bond angle is closer to the value
for a symmetric tetrahedron, even when the liquid density is imposed. 
In terms of simple models of the structure, the AIMD simulation is approximately
consistent with the $\beta$-tin hypothesis \cite{Isherwood1972}.

In conclusion, QENS data constitutes experimental evidence that can
differentiate between different theoretical and numerical models. The {\it
ab initio} model presented here is in very good agreement with all new
and existing, structural and dynamic experimental data. Further details of the liquid structure
are most likely to be obtained from more in-depth investigation of the
MD trajectories.

\begin{acknowledgments}

The authors thank R. Bellissent (CEA, Grenoble) for useful discussions
in designing the sample cell and preparing the experiments and P. Martin
(ILL) for his help in preparing the niobium cell. Some of the calculations
were performed on the CCRT/CEA clusters. Part of this work was supported
by the European Community's Human Potential Program under contract
HPRN-CT-2002-00307, DYGLAGEMEM.
\end{acknowledgments}

\end{document}